\documentclass[prl,aps,floats,twocolumn]{revtex4}
\usepackage{epsfig}
\newcommand{\be}{\begin{equation}}
\newcommand{\ee}{\end{equation}}
\newcommand{\bea}{\begin{eqnarray}}
\newcommand{\eea}{\end{eqnarray}}

\begin{document}

\setlength{\unitlength}{1mm}

\title{On Dark Matter Self-Interactions from Higher
Dimensional Gravity}

\author{Nicolas Chatillon\footnote{nchatill@physics.syr.edu}, Cosmin
Macesanu\footnote{cmacesan@physics.syr.edu}, Aleksandr
Pinzul\footnote{apinzul@physics.syr.edu} and Mark
Trodden\footnote{trodden@physics.syr.edu}}

\affiliation{Department of Physics, Syracuse University,
Syracuse, NY 13244-1130, USA.}

\begin{abstract}
It has recently been suggested that in a brane world scenario with
large extra dimensions, short distance gravitational interactions
can enhance the dark matter scattering cross-section in a velocity
dependent way. Such a modification may then help to address
possible problems with non-interacting cold dark matter on
galactic and sub-galactic scales. We argue that, considering the
singular character of the higher dimensional Newtonian potential,
the scattering cross-section is UV-dependent, depending ultimately
on the underlying quantum gravity theory considered. We
demonstrate that for a wide class of unitary short distance
regularizations, the actual cross-section is velocity-independent
and does not significantly affect dark matter substructure. We
comment on the problem of thermalization of ultra-light cold dark
matter by gravitational interactions in the early universe.
\end{abstract}
\maketitle

The concept of extra spatial dimensions with size much larger than
the Planck scale has attracted considerable interest in recent
years due to its connection with the gauge hierarchy problem and
its experimental testability. If matter is approximately confined
to a $4$-dimensional brane, and only gravity propagates in the
full $(4+n)$-dimensional space-time, compactification radii up to
a fraction of millimeter become possible, in the case where the
fundamental scale of gravity is close to the electroweak scale, as
required to address the gauge hierarchy
problem~\cite{Arkani-Hamed:1998rs}. Alternate solutions
are offered by theories with warped extra dimensions \cite{RS2}.

Models with flat extra dimensions have in common the presence of a
quasi-continuum tower of Kaluza-Klein (KK) gravitons, with masses as
low as an eV, and consequently feature a modified Newtonian potential
at mesoscopic scales.

In a cosmological context, this raises the question of the impact
of enhanced gravitational interactions between objects which are
otherwise weakly interacting, such as dark matter particles.

There is now strong evidence that dark matter comprises
approximately $25\%$ of the energy budget of the universe
\cite{a}, under the assumption that General Relativity itself is
unmodified. In the standard picture dark matter is made of a
neutral, long-lived and weakly interacting elementary particle,
whose exact nature remains to be discovered. However, detailed
comparisons of numerical simulations based on this model
(according to which dark matter halos may have a cuspy core with a
large density gradient \cite{Navarro:1996he}) with recent
astrophysical observations \cite{obs no cusp} (in which this
structure does not seem to be observed), have sparked a debate
about whether modifications of the standard cold dark matter (CDM)
picture are necessary.

Firmani {\it et al.}~\cite{Firmani:2000ce} have suggested that a
velocity-dependent dark matter scattering cross-section could
provide adequate suppression of substructure on the scales of galaxies
and galaxy clusters. These authors suggested the empirical
relation
\begin{equation}
\frac{\sigma_{xx}}{m_x} \approx 4\times 10^{-25}\Big(\frac{100\
km/s}{v}\Big)\ \frac{cm^2}{GeV} \ , \label{empirical
cross-section}
\end{equation}
the relevance of which is still subject to debate (see
e.g.~\cite{Sanchez-Salcedo:2005vc}). Here $m_x$ is the mass of a
CDM particle. Note that the velocity
dispersion for such dark matter systems is $v/c \approx 10^{-3}$.

Qin, Pen and Silk have recently proposed~\cite{Qin:2005pf} that such a
scattering cross-section could find a natural origin in
higher-dimensional gravitational interactions in a brane
world model of the Arkani-Hamed, Dimopoulos and Dvali (ADD) type.
Indeed, below the compactification
radius $R$, the Newtonian potential $-G m^2_x/r$ between two dark
matter particles is modified to
\begin{equation}
V(r) = -\frac{G R^n m_x^2}{r^{1+n}} \label{higher dim potential}
\end{equation}
for $n$ large extra dimensions, where $G$ is the 4D Newton
constant. In the ADD model the radius of the compactified
extra dimensions is
\begin{equation}
R\sim M_D^{-1}\left(\frac{M_P}{M_D}\right)^{2/n} \ ,
\end{equation}
with $M_D$ (the 4+n dimensional Planck mass) taken to be
at the TeV scale. $M_D$ plays the role of a UV
cut-off for the effective field theory description,
 above which quantum gravity effects are expected to become important.

For this potential, the classical
cross-section is then given, for $n\geq 1$, by~\cite{classical
cross-section}
\begin{equation}
\sigma_{xx}= \pi (n+1) (n-1)^{\frac{1-n}{1+n}} \Big(\frac{G R^n
m_x}{2v^2}\Big)^{\frac{2}{n+1}} \ ,
\label{classical capture cross-section}
\end{equation}
where the mean squared relative velocity of the two interacting
particles  is $2v^2$.

Since the de Broglie wavelength for the interacting particles
$\lambda = {2\pi}/{m_x v}$ is larger than the interaction scale
$\sim 1/M_D$, the calculation should be reconsidered in the
quantum regime. Based on the analysis of Vogt and Wannier
\cite{Vogt and Wannier} for $n=3$ and of Joachain \cite{Joachain}
for general $n$, the authors of~\cite{Qin:2005pf} argue that the
result~(\ref{classical capture cross-section}) is unmodified up to
a constant of order of unity. They then fit this expression to
the phenomenological cross-section (\ref{empirical
cross-section}). Fitting the functional form picks out $n=3$ for
the number of extra dimensions, and fitting the magnitude selects a
mass for the dark matter
candidate in the axion range $m_x\sim 3\times 10^{-16}$GeV.

In this letter we consider the general problem of calculating the
corrections to the dark matter gravitational interaction
cross-section due to the presence of extra dimensions. We argue
that, due to the singular character of the potential, the quantum
mechanical elastic scattering cross-section is ambiguous, and that
the assumptions of Vogt and Wannier used to derive the
UV-independent result of order of (\ref{classical capture
cross-section}) do not apply to this problem. We find that, in
general, the cross-section is UV-dependent (see also \cite{KPW}),
and by using several simple regularization procedures, one obtains
\begin{equation}
\sigma_{xx}\sim \frac{m_x^6}{M^8_D} \ ,
\label{simplecrosssection}
\end{equation}
with no dependence on velocity.
This in turn implies that the ADD-type gravitational interaction is
too small to affect the dark matter distribution in galaxies.

For $n\geq 1$, the potential (\ref{higher dim potential}) belongs
to the class of {\it singular potentials}, whose properties are
reviewed extensively in \cite{review singular potentials}.

Classically and in the attractive case, their main property is
that a particle of non-zero angular momentum $L$ may spiral to the
origin. This can be seen from the radial effective potential of
the reduced two-body system including the centrifugal barrier
\begin{equation}
V_{eff}(r)=-\frac{GR^n m_x^2}{r^{1+n}}+\frac{L^2}{2m_x r^2} \ ,
\label{effective potential}
\end{equation}
which is unbounded from below for $r\rightarrow 0$ and $n>1$. The
critical impact parameter below which the incoming particle may
cross the barrier and spiral to $r=0$ (in a finite time) defines
the classical {\it capture} cross-section. Although the particle
reaches the origin with infinite speed, the scattering angle
integral
\begin{equation}
\theta_{\rm scatt}=\pi-2\int_{r_{min}}^\infty
\frac{dr}{r^{2}}
\frac{L}{\sqrt{2 m_x}}\Big[E-V_{eff}(r)\Big]^{-1/2}
\label{scattering angle integral}
\end{equation}
is finite when the minimum approach distance $r_{min}\rightarrow
0$, as it is the case here. This implies the existence of a
tangent to the trajectory at the origin, and the outgoing orbit
may be connected to the infalling one in a unique way by assuming
conservation of energy and angular momentum through $r=0$. Note
that, without these assumptions, the classical trajectory already
presents some ambiguity in connecting ingoing and outgoing orbits.
In any case, because the spiralling trajectories induce random
large angle scattering, while the deflection of non captured
particles is comparatively small, the capture cross-section
is a good approximation to the total transport
cross-section
\begin{equation}
\sigma_t = \int (1-\cos \theta)d\sigma \ ,
\end{equation}
in the classical case.

As apparent from (\ref{effective potential}), the case $n=1$ has a
transition behavior : the spiralling trajectories are possible
only for small enough angular momentum $L<(2G R m_x^3)^{1/2}$.
In that case, the scattering integral (\ref{scattering angle integral}) is infinite,
meaning that there is no tangent at the origin and
thus no natural unambiguous prescription for the outgoing
trajectory.

In the quantum mechanical case, there is no longer a clear
separation between the captured and non-captured trajectories, as
the wave function probes all impact parameters (or, stated
differently, tunneling occurs through the centrifugal barrier for
arbitrarily small positive energies). In contrast to the
non-singular case, in which square integrability of the wave
function is sufficient to determine the boundary condition at the
origin, there are now two {\it a priori} admissible wave function
behaviors near $r=0$; an ingoing one $\psi_{\rm in}(r)$ and an
outgoing one $\psi_{\rm out}(r)$, which infinitely rapidly
oscillate at the origin
\begin{equation}
\psi_{\rm in/out}(r) \sim r^{\frac{n-3}{4}}\exp\left(\pm \frac{2
i}{n-1}\sqrt{\frac{G R^n m_x^3}{r^{n-1}}} \right) \ . \label{wave
functions near r=0}
\end{equation}
(We consider only the s-wave case for this low energy
scattering problem).

This reflects the classical connection ambiguity in the quantum
case. The problem can be traced back to the fact that the UV
region affects the wavefunction  much more than in the case of
regular potentials. While very short distance modifications of
such potentials would not dramatically alter low-energy
scattering, the missing boundary condition for singular potentials
demonstrates a form of UV-sensitivity. This boundary condition may
be understood as arising, for example, from a short-distance
regular modification of the potential, thus depending on the
specific choice of regularization.

The possible boundary conditions may be reduced to a smaller
subset by requiring hermiticity of the Hamiltonian, to ensure that
elastic scattering is actually a unitary process. This implies
orthogonality of the wave functions, which in turn imposes at
$r=0$ linear combinations of the form
\begin{equation}
\psi(r) \sim e^{i\theta} \psi_{\rm in}(r) + e^{-i\theta}\psi_{\rm
out}(r) \ ,
\end{equation}
with the same relative phase factor $e^{i\theta}$ for all
eigenfunctions \cite{Case:1950an}, up to an overall normalization
coefficient. Note that hermiticity does not
fix the phase factor
$e^{i\theta}$, which may be obtained from experiment or derived
in a particular regularization.

Vogt and Wannier \cite{Vogt and Wannier} have studied the $-1/r^4$
potential in the context of the scattering of ions by polarization
forces, but their approach may be applied for general $n$. They
observed that the wave functions (\ref{wave functions near r=0})
oscillate extremely rapidly for
\begin{equation}
r < (G R^n m_x^3)^{\frac{1}{n-1}}
=\frac{1}{M_D}\left(\frac{m_x}{M_D}\right)^{\frac{3}{n-1}} \ .
\label{oscillation radius}
\end{equation}
If the regularization of the potential occurs at a length scale smaller than this value, a
small energy spread in the incoming wave will dramatically affect
the scattering phases, making the outgoing wave incoherent. Thus,
in practice, a sink boundary condition, $\psi_{\rm out}(r)=0$, at $r=0$
correctly describes scattering (although note that an erroneous
interpretation of this condition as fundamental would lead to the
conclusion that the process is non unitary). In this case the
dominant contribution to the cross-section is the capture one, as
defined by the ratio of the ingoing flux at the origin to the
incoming one from infinity. Then the quantum cross-section is of the same order of
magnitude as the classical capture one~\cite{Vogt and Wannier}; this is the result
(\ref{classical capture cross-section}) used in~\cite{Qin:2005pf}.

However, when the cut-off length-scale is larger than
(\ref{oscillation radius}), the wave function does not experience
the rapid oscillations required to avoid the large distance
effects of a specific regularization, and the results become
UV-dependent. In particular, resonant scattering,
drastically differing from~(\ref{classical capture
cross-section}), may occur.

Returning now to the specific case of brane-world gravity, the fundamental
Planck length
$1/M_D$ provides a natural cut-off. Assuming the dark matter
mass $m_x < M_D$, one obtains
\begin{equation}
\frac{1}{M_D} >
\frac{1}{M_D}\left(\frac{m_x}{M_D}\right)^{\frac{3}{n-1}} \ ,
\end{equation}
which implies that wavefunctions do not have room to oscillate
rapidly before reaching the region in which quantum gravity
corrections modify the theory. The scattering cross-section is
thus regularization-dependent and the effective sink boundary
condition used in~\cite{Qin:2005pf} is in general not sufficient to
describe the process.

%%%%%%%%%%%%%%%

To investigate possible outcomes, we consider some simple test regularizations,
effective below the cutoff $M_D^{-1}$. For instance,
consider the potential resulting from truncating the tower of
KK gravitons at the scale $M_D$
\begin{equation}
V(r) = -G R^n m_x^2 \int_{0}^{M_D} \rho(m) \frac{e^{-mr}}{r} dm \ ,
\label{reg_v}
\end{equation}
where the density of KK modes is
\begin{equation}
\rho(m)=\frac{2\pi^{n/2}}{\Gamma(n/2)} m^{n-1} \ .
\end{equation}
This potential behaves as $-G R^{n} m_x^2/r^{n+1}$ for $R \gg r\gg M_D^{-1}$,
and as $-G' m_x^2/r$ for $r \ll M_D^{-1}$ (with $G' \sim
1/M_D^2$, such that the potential is continuous).

The resulting potential now has a manageable $1/r$ behavior at
scales $\ll 1/M_D$. Using standard phase shift analysis, one
obtains the $s$-wave scattering cross-section~(\ref{simplecrosssection}),
which is negligibly small and velocity-independent.

This result can be checked by other methods. Assuming a
short-distance potential at best as steep as $r^{-1}$, the s-wave
cross-section may be calculated using the Born approximation
$$
\sigma^{Born}_{l=0} \approx 16 \pi m_x^2 \left[\int_0^\infty dr\
r^2 J_0^2(m_x v_\infty r) V(r)\right]^2 \ ,
$$
where $J_0$ is the corresponding Bessel function. For $n\geq 3$, the
``KK-regularized" potential above yields
\begin{equation}
\sigma^{Born}_{l=0} \sim \frac{m_x^6}{M_D^8}
\left[1+O\left(\frac{m_x v}{M_D}\right)\right] \ (n \geq 3) \ .
\end{equation}

As another test of regularization dependence we may use
the exactly solvable (for $s$-wave) potentials
\begin{equation}
V(r)= -\frac{G R^3 m_x^2}{(r^2+M_D^{-2})^2} {\rm \ or \ } -\frac{G
R^3 m_x^2}{r^4+M_D^{-4}} \ ,
\end{equation}
for $n=3$, which are finite at the origin. In either case the result is similar to
(\ref{simplecrosssection}).

%%%%%%%%%%%%%%%%

>From the above reasoning, in particular using the Born
approximation for $n>2$, we see that any short distance
regularization not strictly steeper than $r^{-1}$ will not provide
a contribution to the cross-section significantly larger than the
$r^{-(n+1)}$ tail. This establishes the generality of the result
(\ref{simplecrosssection}) under these assumptions. It
should be noted, however, that steeper regularizations may
strongly alter this result and enhance the cross-section: the
simplest example of an infinite potential wall at $r=M_D^{-1}$
would result in hard sphere scattering with $\sigma_{xx}\sim
M_D^{-2}\gg m_x^6/M_D^{8}$.

It is instructive to compare this result to the tree-level
relativistic scattering cross-section due to the exchange of a
tower of KK gravitons. A similar phenomenon of
UV-dependence appears in the ADD model as the sum over
KK propagators diverges for $n \geq 2$ \cite{sum KK
divergent in ADD}. This is regularized by limiting the sum to
gravitons with mass less than the $M_D$ scale (as in Eq.
(\ref{reg_v})). The $v\ll c$ limit obtained after cutting off the
KK sum at $M_D$ again reproduces (\ref{simplecrosssection}).

Higher-dimensional gravity being in itself non-singular (when
matter is also higher-dimensional), this divergence and
with it the UV-dependence that we encountered in the
non-relativistic case, can be attributed to the localization of
matter on an infinitely thin brane. A natural regularization would
take into account the finite width of the brane of order
$M_D^{-1}$, and provide an unambiguous answer to the scattering
problem without detailed knowledge of the underlying quantum
gravity theory.

%%%%%%%%%%%%%%%%%%%%%%%%%

The early universe evolution of ultra-light particles in the
presence of enhanced short distance gravity has also been
discussed recently \cite{KPW}, with the worry that they may be
thermalized in the radiation era and disqualified as cold dark
matter candidates.
However, the rate at which the light particles gain energy
through collisions with the thermalized matter  is proportional to
the  relativistic cross-section interaction, which,
as discussed above, is suppressed by
eight powers of $M_D$; thus the effect is not significant at
temperatures  below the $M_D$ scale.
The same  conclusion has been reached in \cite{KPW}, once
quatum effects are taken into account.

%%%%%%%%%%%%%%%%%%%%%%%%%

In this letter, we have studied the impact of boundary conditions
at the origin on the elastic scattering of dark matter particles
by singular brane world gravitational interactions. We have shown
that the cut-off length $M_D^{-1}$ is too large for the sink
prescription of ~\cite{Vogt and Wannier} to apply. For
sufficiently shallow regularizations, we obtain a cross-section of
order of $\sigma_{xx}\sim M_D^{-2 }(m_x/M_D)^6$, which is much
smaller than $\sigma_{xx}\sim \frac{c}{v}M_D^{-2}(m_x/M_D)^{1/2}$,
which has been proposed to help to reconcile recent observations
with simulations of dark matter systems and without the requisite
velocity dependence. It may be interesting to consider in more
detail the case of dark matter particles with masses close to the
cutoff, for which this suppression would be much less important.
Finally, we have discussed the thermalization  of
ultra-light particles gravitationally interacting in the early
universe, with the conclusion that they will remain cold if
produced below the $M_D$ scale.

{\it Acknowledgments} NC and MT are supported by the NSF under
grant PHY-0354990 and by Research Corporation. NC, CM and AP are
supported by DOE grant DE-FG02-85ER40231.

\end{document}